\documentclass{nature}
\usepackage{amssymb}
\usepackage{rotating}
\usepackage{amssymb}
\usepackage{graphicx}
\usepackage{subfigure}
\usepackage{algorithm}
\usepackage{algorithmic}
\usepackage{epsfig}
\usepackage{comment}
\usepackage{url}
\usepackage{float}
\usepackage{slashbox}
\usepackage{enumerate}
\usepackage[margin=1in]{geometry}
\usepackage[T1]{fontenc}

\newcommand{\tabincell}[2]{\begin{tabular}{@{}#1@{}}#2\end{tabular}}

\title{Enhanced Community Structure Detection in Complex Networks with Partial Background Information}

\author{Zhong-Yuan Zhang$^{1\ast}$, Kai-Di Sun$^{1}$, Si-Qi Wang$^{1}$ \\ $^\ast$ To whom correspondence should be addressed. \\ Email: zhyuanzh@gmail.com}
\date{}
\begin{document}

\maketitle
\begin{affiliations}
 \item School of Statistics, Central University of Finance and Economics, P.R.China
\end{affiliations}
\begin{abstract}
Community structure detection in complex networks is important since it can help better understand the network topology and how the network works. However, there is still not a clear and widely-accepted definition of community structure, and in practice, different models may give very different results of communities, making it hard to explain the results. In this paper, different from the traditional methodologies, we design an enhanced semi-supervised learning framework for community detection, which can effectively incorporate the available prior information to guide the detection process and can make the results more explainable. By logical inference, the prior information is more fully utilized. The experiments on both the synthetic and the real-world networks confirm the effectiveness of the framework.
\end{abstract}

\maketitle

\section*{Introduction}
\label{intro}
Community structure detection in complex networks is of critical importance for understanding not only the network topology, but also how the network works \cite{Girvan02, newman2004finding, newman2011}. In many real applications,  the revealed communities often correspond to functional modules of the network, such as pathways in metabolic networks, or a group of people that have common interest. These functional modules can be considered building blocks of networks. Furthermore, dynamics in the networks with communities can be very different from those without communities.

However, it is very hard to give a general and widely-accepted definition of community structure due to the complexity of real problems, and most of the revealed communities are model-based, which makes the results hard to explain, or in other words, the correctness and meanings of the communities cannot be confirmed without the background information about the functions of the nodes. Hence if the background information can be effectively incorporated to guide the process of community structure detection, we will get much better results. In our previous work, we have proposed a semi-supervised learning framework to incorporate
two types of background information into community detection: \emph{must-link} and \emph{cannot link} \cite{semi}. An
interesting question is: \emph{How to make better use of the prior knowledge to do more with less?}

In this paper, based on the results of [4] and logical inference, we propose an enhanced semi-supervised learning framework for community structure detection, which can incorporate the available prior information more effectively. The most important contribution of the framework is that it adds a logical inference step to more fully utilize the two types of prior information, must-link constraints and cannot-link constraints, and can more effectively combine the information with the topology structure of networks to guide the detecting process. The experimental results show that the proposed method can significantly improve the detection performance. We also evaluate which type of constraints is more useful, indicating that the constraints of must-link contribute much more than that of cannot-link.

\section*{Results}
\label{results}
In this section, we empirically tested the effectiveness of the enhanced semi-supervised learning framework for community detection.
To do this, we applied NMF, spectral clustering and InfoMap with the revised adjacency matrices to several well-studied networks.
\subsection{Data Description}\label{synthetic}
Both the synthetic and the real-world networks were used in our experiments. Details are as follows:
\begin{enumerate}[1)]
\item GN \cite{Girvan02}: The GN network, also known as the ``four groups'' network,  has $128$ nodes which are divided into four equalized sized non-overlapping communities with 32 nodes each. On average, each node has $Z_{in}+Z_{out}=16$ neighbors, or in other words, it randomly connects with $Z_{in}$ nodes in its own community and $Z_{out}$ nodes in other communities. As expected, with an increasing $Z_{out}$, the community structure will become less and less clear and the problem more challenging. In our experiment, we set $Z_{out}$ to 10 and $Z_{in}$ to 6.
 \item LFR \cite{LFR}:
     Compared with GN networks, LFR networks are more realistic. In LFR networks, both the degree and the community size distributions obey power laws, with exponents $\gamma$
and $\beta$. Each node has a fraction $1-\mu$ of its neighbors in its own community and a fraction $\mu$ in other communities. Here $\mu$ is called the mixing parameter.

We set the parameters of the LFR network as follows: the number of nodes was $1000$, the average degree of the nodes was 20, the maximum degree was 50, the exponent of degree distribution $\gamma$ was 2 and that of community size distribution $\beta$ was 1, and the mixing parameter $\mu$ was  $0.9$. The communities did not overlap with each other.
\item  Football team network \cite{Girvan02}: this dataset is about the network of American football games (not soccer) between
115 college teams during regular season Fall 2000. The edges connect the teams that had games.
\end{enumerate}
\subsection{Assessment Standards}
\label{measure}
In this submission, we used the normalized mutual information (NMI)\cite{strehl2003cluster} to assess the effectiveness of our approach. The value can be formulated as follows:
$$\hspace{-2.2mm}
  \mbox{NMI}(M_1, M_2) = \frac{\sum\limits_{i=1}^{k}
     \sum\limits_{j=1}^{k}n_{ij}\log\displaystyle\frac{n_{ij}n}{n_i^{(1)}n_j^{(2)}}}
  {\sqrt{\left(\sum\limits_{i=1}^{k}n_i^{(1)}\log\displaystyle\frac{n_i^{(1)}}{n}\right)
  \left(\sum\limits_{j=1}^{k}n_j^{(2)}\log\displaystyle\frac{n_j^{(2)}}{n}\right)}},
$$
where $M_1$ is the ground-truth community label and $M_2$ is the computed community label, $k$ is the community number, $n$ is the number of nodes, $n_{ij}$ is the number of nodes in the ground-truth community $i$ that are assigned to the computed community $j$, $n_i^{(1)}$ is the number of nodes in the ground-truth community $i$ and $n_j^{(2)}$ is the number of nodes in the computed community $j$, $\log$ is the natural logarithm.

In general, the higher the NMI, the better the result.
\subsection{Results Analysis}
\label{simulation}
Consequently, we compared the results of NMI obtained by the models with and without prior information. For an undirected network
with $n$ nodes, there are totally $n(n-1)/2$ node pairs
available. We randomly selected a percentage of node pairs,
and determined whether the pairs were ML or CL. Then we incorporated these pairwise constraints into the adjacency matrix $A$ to get $B^{[1]}$ and $B^{[2]}$. To evaluate which type of constraints is more useful, we further used the following three matrices, where only one type of constraints was incorporated:
\begin{equation}\label{eq:03}
B^{[1]\_{\footnotesize ML}}_{ij} = \left\{\begin{array}{rcl}
         B^{[1]}, & & \mbox{if}\ i\ \mbox{\&}\ j\ \mbox{are known to be ML}\\
         A_{ij}, & & \mbox{otherwise},
      \end{array}\right.
\end{equation}
\begin{equation}\label{eq:04}
B^{[1]\_{\footnotesize CL}}_{ij} = \left\{\begin{array}{rcl}
         B^{[1]}, & & \mbox{if}\ i\ \mbox{\&}\ j\ \mbox{are known to be CL}\\
         A_{ij}, & & \mbox{otherwise},
      \end{array}\right.
\end{equation}
and based on $B^{[1]\_{\footnotesize ML}}$,
\begin{equation}\label{eq:03}
B^{[2]\_{\footnotesize ML}}_{tk} = \left\{\begin{array}{rcl}
         \alpha, & & \mbox{if}\ i\ \mbox{\&}\ t\ \mbox{are ML,}\\
                 & & \mbox{and}\ i\ \mbox{\&}\ k\ \mbox{are ML}\\
         B^{[1]\_{\footnotesize ML}}_{tk}, & & \mbox{otherwise},
      \end{array}\right.
\end{equation}
Note that if we only consider the constraints of CL, the information cannot be enhanced, so $B^{[2]\_{\footnotesize CL}}$ should be identical with $B^{[1]\_{\footnotesize CL}}$, and was not defined again.
 The
results were averages of ten trails and have summarized in Fig. 1 and 2. From these figures, one can observe that:
(1) The averaged NMI of the semi-supervised learning with and without the information-enhanced step increases with the increasing percentage of ML and CL pairs constrained; (2) The information-enhanced step does significantly increase the detection performance. For example, given $5$ percent of pairs constrained on GN networks, the NMI of the non-enhanced semi-supervised NMF ($=55.78\%$) is forty percent higher than that of the unsupervised one ($=10.51\%$), and the NMI of the enhanced NMF ($=87.35\%$) is further improved by more than thirty percent; 3) For the GN networks, the spectral clustering is slightly better, while for the LFR networks, the InfoMap is better; 4) The type of ML constraints contributes much more than that of CL constraints. Note that for the LFR benchmarks, the results of NMF on $B^{[2]\_{\footnotesize ML}}$ are slightly better because, for $B^{[2]}$, with a high percentage of pairs constrained, many elements are replaced by zero due to CL constraints, and some columns only have few non-zero elements, reducing the clustering performance.

\begin{figure}
\centering
\includegraphics[height=160mm,width=70mm]{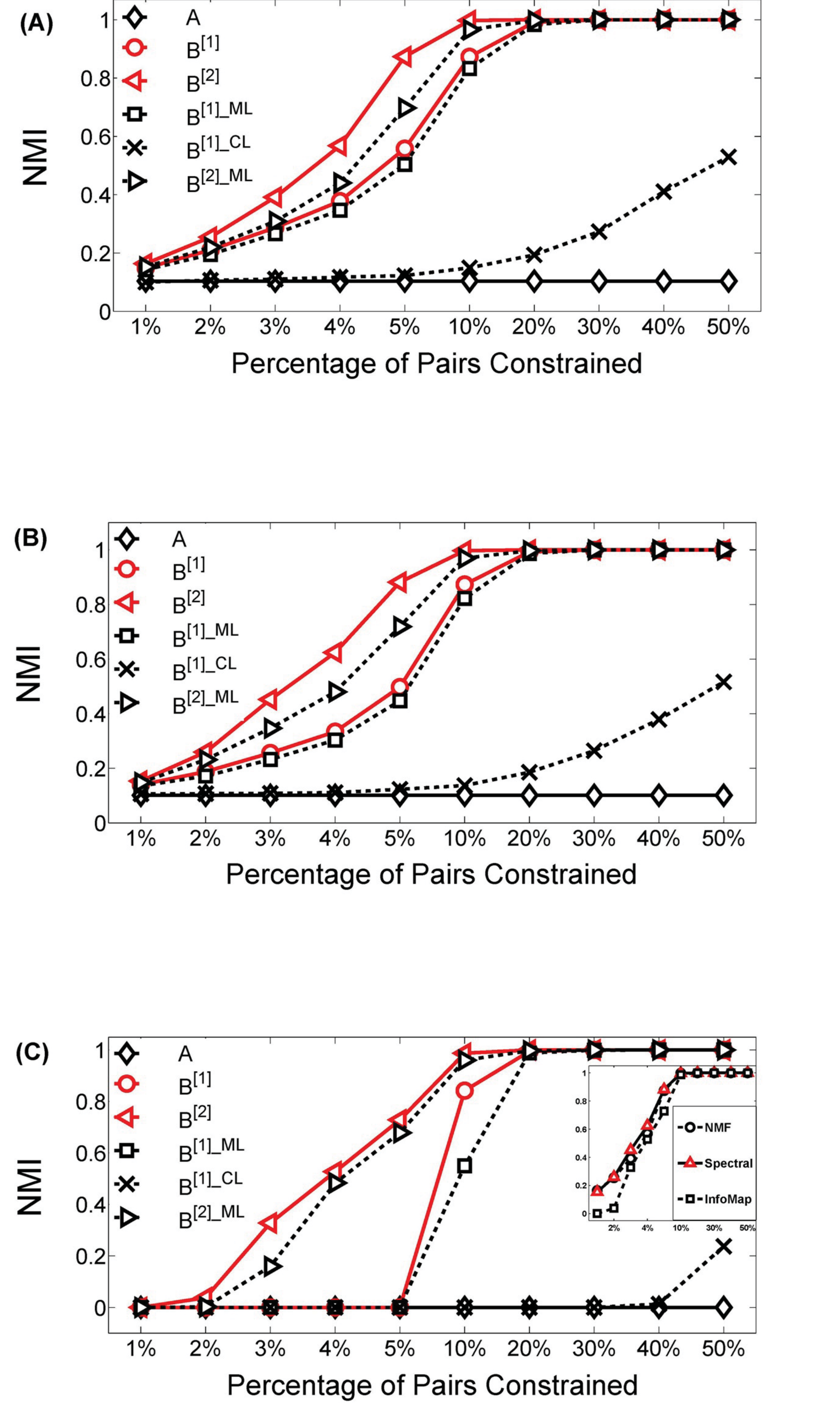}
 \caption{Averaged NMI of (A) NMF, (B) spectral clustering and (C) InfoMap under different percentage of node pairs constrained on GN benchmarks.  The legend denotes the results on different objective matrices. The inset compares the NMI results obtained by NMF, spectral clustering and InfoMap on $B^{[2]}$.}\label{Fig:03}
\end{figure}

\begin{figure}
\centering
\includegraphics[height=160mm,width=70mm]{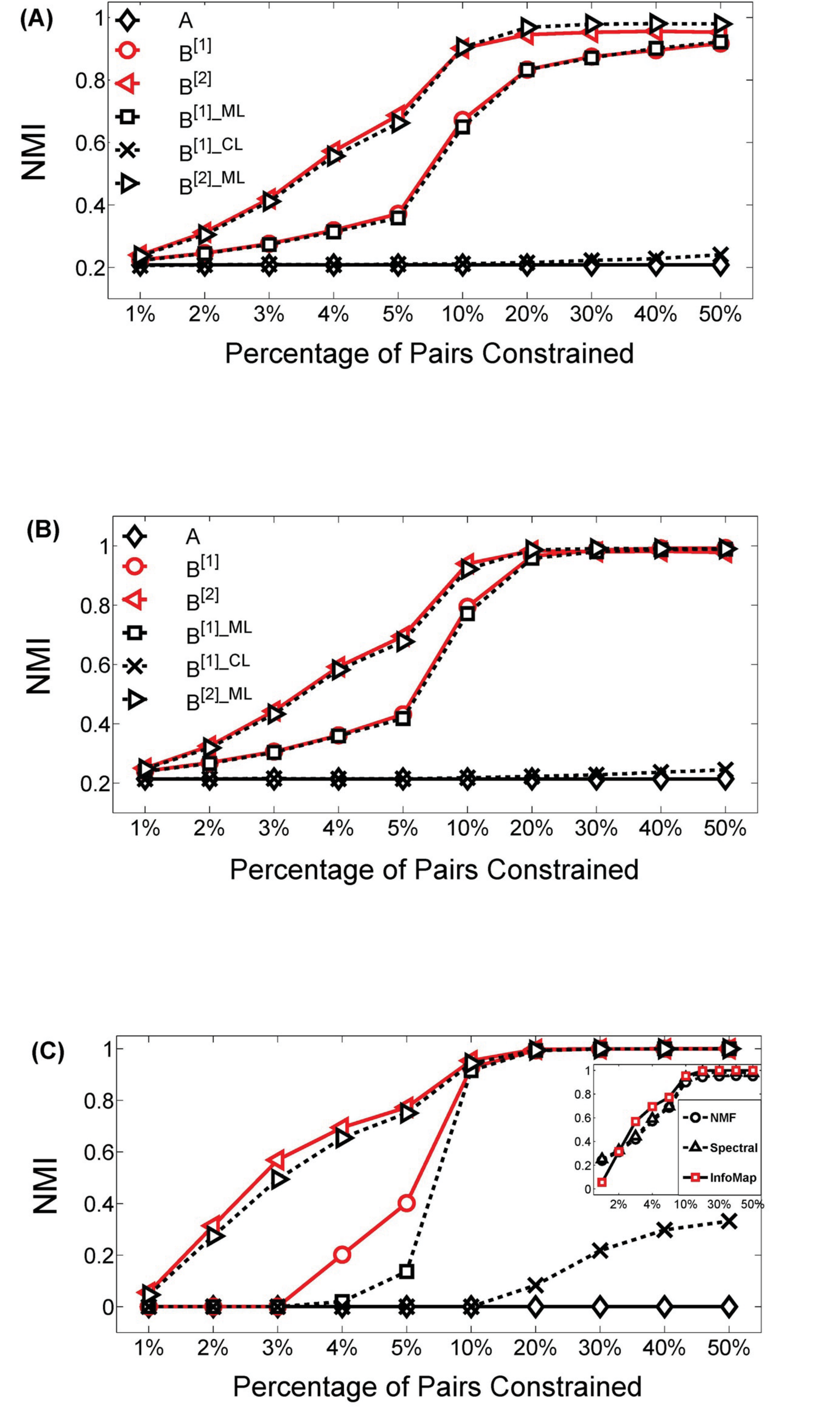}
\caption{Averaged NMI of (A) NMF, (B) spectral clustering and (C) InfoMap under different percentage of node pairs constrained on LFR benchmarks.  Meanings of $A$, $B^{[1]}$, $B^{[2]}$, $B^{[1]\_{\footnotesize ML}}$, $B^{[1]\_{\footnotesize CL}}$ and $B^{[2]\_{\footnotesize ML}}$ are identical with that in Fig.\ref{Fig:03}. The inset compares the NMI results obtained by NMF, spectral clustering and InfoMap on $B^{[2]}$.}\label{Fig:04}
\end{figure}
\subsection{A Case Study: College Football Network}
\label{case}
In this subsection, we analyzed the football team network as a case study to show the effectiveness of our semi-supervised learning framework.

In the football network, there are 115 nodes (teams), and they belong to 12 different conferences. Most of them played against the ones in the same conference
more frequently. However, there are also abnormal teams that played more frequently against the ones in other conferences, including the teams 37, 43, 81, 83, 91 (in conference IA
Independents), 12, 25, 51, 60, 64, 70, 98 (in conference Sunbelt), 111, 29 and
59. For more details, please see [4].

Combined with our previous work \cite{semi}, we set the community number to 11, and the teams 37, 43, 81, 83, 91 (conference IA
Independents) did not have ground-truth conference labels, or in other words, there were 110 labeled teams and $110\times(110-1)/2=5995$ team pairs available. Firstly, we randomly selected some pairs in them as prior information: if the two teams were in the same
conference, they were must-link (ML), otherwise, they were cannot-link (CL), and then we proceeded to implement the information-enhanced step. Finally, we applied NMF on the revised adjacency matrices to give the partitioning results.

Fig. 3 gives the partitioning results of NMF corresponding to different percentage of pairs constrained (with and without information-enhanced step), from which, one can observe that:
\begin{enumerate}[i)]
\item When no prior information is given, there are five abnormal teams mis-clustered: teams 29, 60, 64, 98, 111.
\item When given $5$ percent of pairs constrained, the number of ML and CL pairs of constrained without the information-enhanced step is 300 (5 percent of node pairs available), and increases to 1130 after the information-enhanced step (18.85 percent of node pairs available). There are three abnormal teams mis-clustered: teams 29, 64, 111. An interesting observation here is that the team 64 is mis-clustered into two different conferences (conference 10 and conference 5) under the non-information-enhanced learning and the information-enhanced learning. The reasons are as follows: in this experiment, before the information-enhanced step, there are 7 CL pairs and no ML pairs related with node 64, among which, two are related with the conference 10 while none is related with the conference 5. After the information-enhanced step, there are 18 CL pairs and no ML pairs related with 64, among which, seven are related with the conference 10 and there are still no pairs related with conference 5. The result is thus guided by the enhanced prior information and the team 64 is reassigned to conference 5.  Note that since the pairs of constrained are selected randomly, another round of the experiment may result in different network partitions. Table \ref{Tab:01} gives more details about the ML and CL pairs related with the team 64.
\item When given $20$ percent of prior knowledge, the number of ML and CL pairs of constrained is 1199 ($20$ percent, before enhancement), and increases to 5651 ($94.26$ percent, after). There is only one team 111 mis-clustered into the conference 12 under the non-enhanced learning. After enhancement,  all the labeled teams are corrected clustered. The teams of the conference 12 are all in the CL set of the team 111, which is very helpful to the community structure detection process. Table \ref{Tab:02} gives more details about the ML and CL pairs related with the team 111. The team 64 is clustered correctly. Before the information-enhanced step, there are 30 CL pairs and 3 ML pairs related with 64, and after enhancement, there are 102 CL pairs (including almost all the labeled teams that are not in conference 11) and 6 ML pairs (including all the labeled teams in conference 11) related with 64.
\end{enumerate}

\begin{table}
\caption{Must link and cannot link pairs of teams related with team 64 and the teams in conference 10 and 5. The boxed nodes are that included in ML or CL.}
\centering\scriptsize
\begin{tabular}{c || c | c}\hline\hline
 Node 64  & ML & CL\\\hline
 Non-enhanced & none & 21, 44, 57, 59, 72, 85, 107\\\hline
 Enhanced & none &  \tabincell{l}{19, 21, 27, 39, 44, 55, 57, 59, 62,\\ 66, 72, 85, 86, 88, 96, 97, 107, 114}  \\\hline
 Conference 10 & \multicolumn{2}{c}{\tabincell{l}{18, \fbox{21}, 28, \fbox{57}, 63, \fbox{66},\\ 71, 77, \fbox{88}, \fbox{96}, \fbox{97}, \fbox{114}}}\\\hline
 Conference 5 & \multicolumn{2}{c}{45, 49, 58, 67, 76, 87, 92, 93, 111, 113}\\
 \hline\hline
\end{tabular}\label{Tab:01}
\end{table}

\begin{table}
\caption{Must link and cannot link pairs of teams related with team 111 and the teams in conference 5 and 12. The boxed nodes are that included in ML or CL.}
\centering\scriptsize
\begin{tabular}{c || c | c}\hline\hline
 Node 111  & ML & CL\\\hline
 Non-enhanced & 45, 67 & \tabincell{l}{3, 7, 10, 13, 15, 18, 24, \\ 26, 27, 46, 79, 82, 88, 101}\\\hline
 Enhanced & 45, 67 & a total of 95 teams\\\hline
 Conference 5 & \multicolumn{2}{c}{\tabincell{l}{\fbox{45}, 49, 58, \fbox{67}, \\ 76, 87, 92, 93, 111, 113}}\\\hline
 Conference 12 & \multicolumn{2}{c}{\tabincell{l}{\fbox{29}, \fbox{47}, \fbox{50}, \fbox{54}, \fbox{59},\\ \fbox{68}, \fbox{74}, \fbox{84}, \fbox{89}, \fbox{115}}}\\
 \hline\hline
\end{tabular}\label{Tab:02}
\end{table}
In summary, our semi-supervised learning framework does make better use of the prior information and can significantly improve the model performance.

\begin{figure}
\centering
\vspace{-25mm}
\includegraphics[height=170mm,width=150mm]{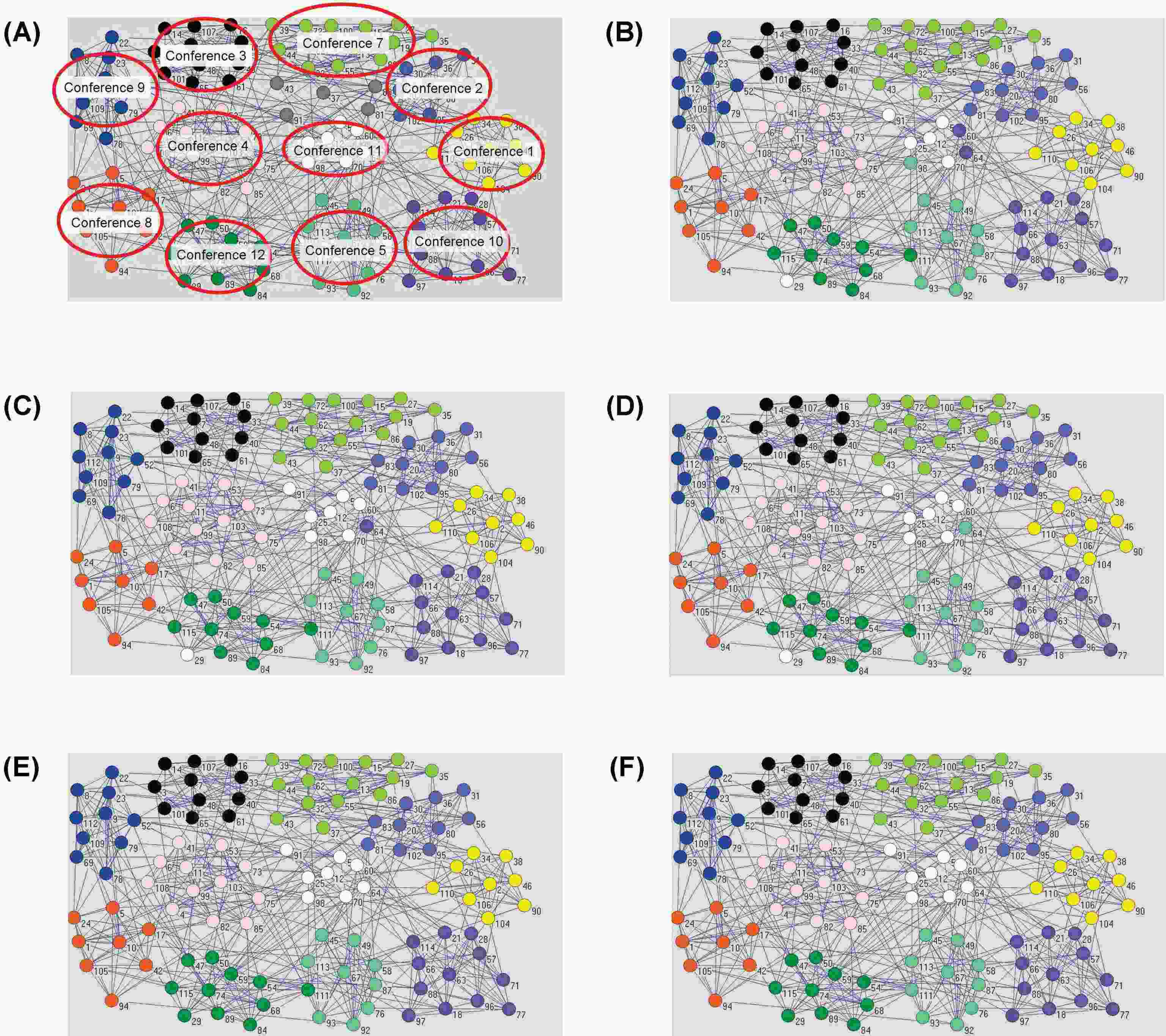}
\caption{Comparison of the semi-supervised learning results of NMF with and without the information-enhanced step corresponding to different percentage of pairs constrained (color online). (a): Real grouping in football dataset. There are 12 conferences of 8-12 teams (nodes) each. Teams in conference 6 are not labeled. (b): Result of NMF without any prior information. (c): Result of NMF given $5$ percent of pairs constrained (without the information-enhanced step). (d): Result of NMF given $5$ percent of pairs constrained (with). (e): Result of NMF given $20$ percent of pairs constrained (without). (f): Result of NMF given $20$ percent of pairs constrained (with), and all the labeled teams are corrected clustered.}\label{Fig:05}
\end{figure}
\section*{Discussion}\label{conclusion}
In this paper, we propose an enhanced semi-supervised learning framework for community structure detection. The framework can add the supervision of must-link (ML) and cannot-link (CL) pairwise constraints into the adjacency matrix. Through the information enhancement based on logical inference, the detection performance can be improved significantly. Note that this step is only feasible under the case of non-overlapping communities. If otherwise, for example, node $i$ has multiple community labels, the prior information that nodes $i$ and $t$ are ML and nodes $i$ and $k$ are CL may not necessarily result in the fact that nodes $t$ and $k$ are also CL.

In addition, we evaluate the contributions of the two types of constraints, indicating that the type of ML constraints is much more important than CL, which is interesting, because in many real scenarios, it is easier to get the constraints of ML (positive constraints) \cite{ML1,ML2}.

An interesting problem which is related with our work is the analysis of dynamic networks, such as detecting the communities in a series of time-varying networks.
 Given the network structure at time $t$, we can find some conservative relationships between nodes and use them as ML and CL constraints to detect the communities in the new network at time $t+1$, which is termed by us as online semi-supervised learning.
 \section*{Methods}
 \subsection{Enhanced semi-supervised learning for community structure detection}
\label{semi0}
In this section, we give our enhanced learning framework for community structure detection. Firstly, given an undirected and unweighted simple graph $G$, we can define the associated symmetric adjacency matrix $A$ as follows:
$$A_{ij} = \left\{\begin{array}{rcl}
         1, & & \mbox{if}\ i\sim j,\ \mbox{or}\ i=j,\\
         0, & & \mbox{if}\ i\nsim j,\ \mbox{and}\ i\neq j,\\
      \end{array}\right.
$$
where $i\sim j$ means there is an edge between nodes $i$ and $j$, and $i\nsim j$ means there is no edge between them.

In many real applications, there is often some prior information available, 
We can try to incorporate this information into the community detection process to make the result more explainable and clear.
Specifically, if we have known that some pairs of nodes are must-link (the two nodes are in the same community, ML), or some pairs of nodes are cannot-link (the two nodes are not in the same community, CL), or both, we can incorporate these pairwise constraints into the adjacency matrix $A$ to get a new matrix $B^{[1]}$ as follows \cite{semi}:
\begin{equation}\label{eq:02}
B^{[1]}_{ij} = \left\{\begin{array}{rcl}
         \alpha, & & \mbox{if}\ i\ \mbox{and}\ j\ \mbox{are known to be ML}\\
         0, & & \mbox{if}\ i\ \mbox{and}\ j\ \mbox{are known to be CL}\\
         A_{ij} & & \mbox{otherwise},
      \end{array}\right.
\end{equation}
In addition, based on logical inferences, one can get further knowledge of the constraints that i) if nodes $i$ and $t$ are ML, and nodes $i$ and $k$ are ML, then $t$ and $k$ should also be ML (\emph{The friend of my friend is my friend}); ii) if nodes $i$ and $t$ are ML, and nodes $i$ and $k$ are CL, then $t$ and $k$ should also be CL (\emph{The friend of my enemy is my enemy}), which leads to the following revision of $B^{[1]}$:
\begin{equation}\label{eq:02}
B^{[2]}_{tk} = \left\{\begin{array}{rcl}
         \alpha, & & \mbox{if}\ i\ \mbox{\&}\ t\ \mbox{are ML,} \ \mbox{and}\ i\ \mbox{\&}\ k\ \mbox{are ML}\\
         0, & & \mbox{if}\ i\ \mbox{\&}\ t\ \mbox{are ML, and}\ i\ \mbox{\&}\ k\ \mbox{are CL}\\
         B^{[1]}_{tk} & & \mbox{otherwise},
      \end{array}\right.
\end{equation}
In this paper, we call this logical inference step as \textbf{information-enhanced step}. After this step, the prior knowledge is more fully utilized. We set $\alpha$ to 2 in this work \cite{semi}.

Note that the idea of the information-enhanced step can be traced back to the balance theory in psychology \cite{balance1,balance}, which was used for explanation of attitude change.
\subsection{Illustrative examples}\label{example1}
We use a small example to intuitively show the ideas behind our framework. Figure 4 (A) is part of a network, where nodes $1, 2, 3, 4, 5, 6$ are in the same community, and node $7$ in another community. Based on some domain knowledge, we may know that nodes $3$ and $5$, nodes $4$ and $5$ are must-link, and nodes $3$ and $7$ are cannot-link. By logical inference, nodes $3$ and $4$ should also be must-link, consequently, nodes $4$ and $7$ should be cannot-link, and finally, nodes $5$ and $7$ should also be cannot-link. The background information is thus significantly enhanced.
\begin{figure}
\includegraphics[height=45mm,width=150mm]{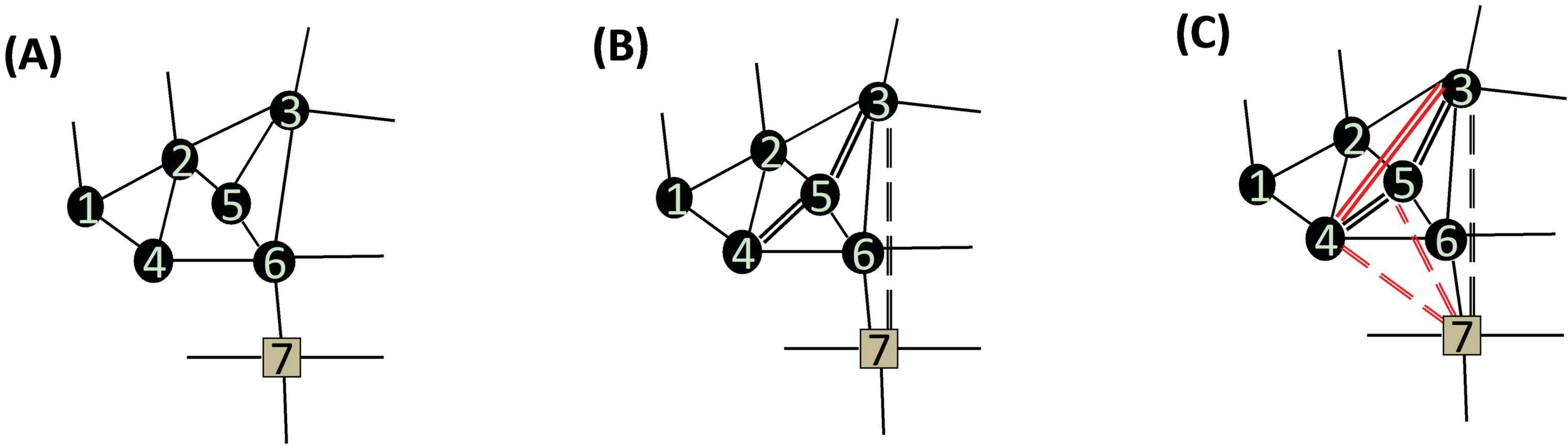}
 \caption{An illustrative example to show the ideas behind our framework.}\label{Fig:06}
\end{figure}

In addition, we illustrate the effectiveness of the approach with a GN network: given an undirected and unweighted ``equally sized four groups'' network with 128 nodes, we want to detect the communities in it. The heatmap of the associated adjacency matrix $A$ is shown as the upper left in Fig. 5. Suppose that we have some prior information about the functions of the nodes, and can thus determine a percentage of pairs of nodes as must-link (ML) or cannot link (CL). These pieces of information on ML and CL are incorporated into the adjacency matrix $A$. As expected, the communities become more and more clear as the percentage of pairs constrained increases. Furthermore, an surprising observation is the effectiveness of logical inferences, which has dramatically improved the data quality (the second row in Fig. 5).
\begin{figure}
\centering
\includegraphics[height=50mm,width=80mm]{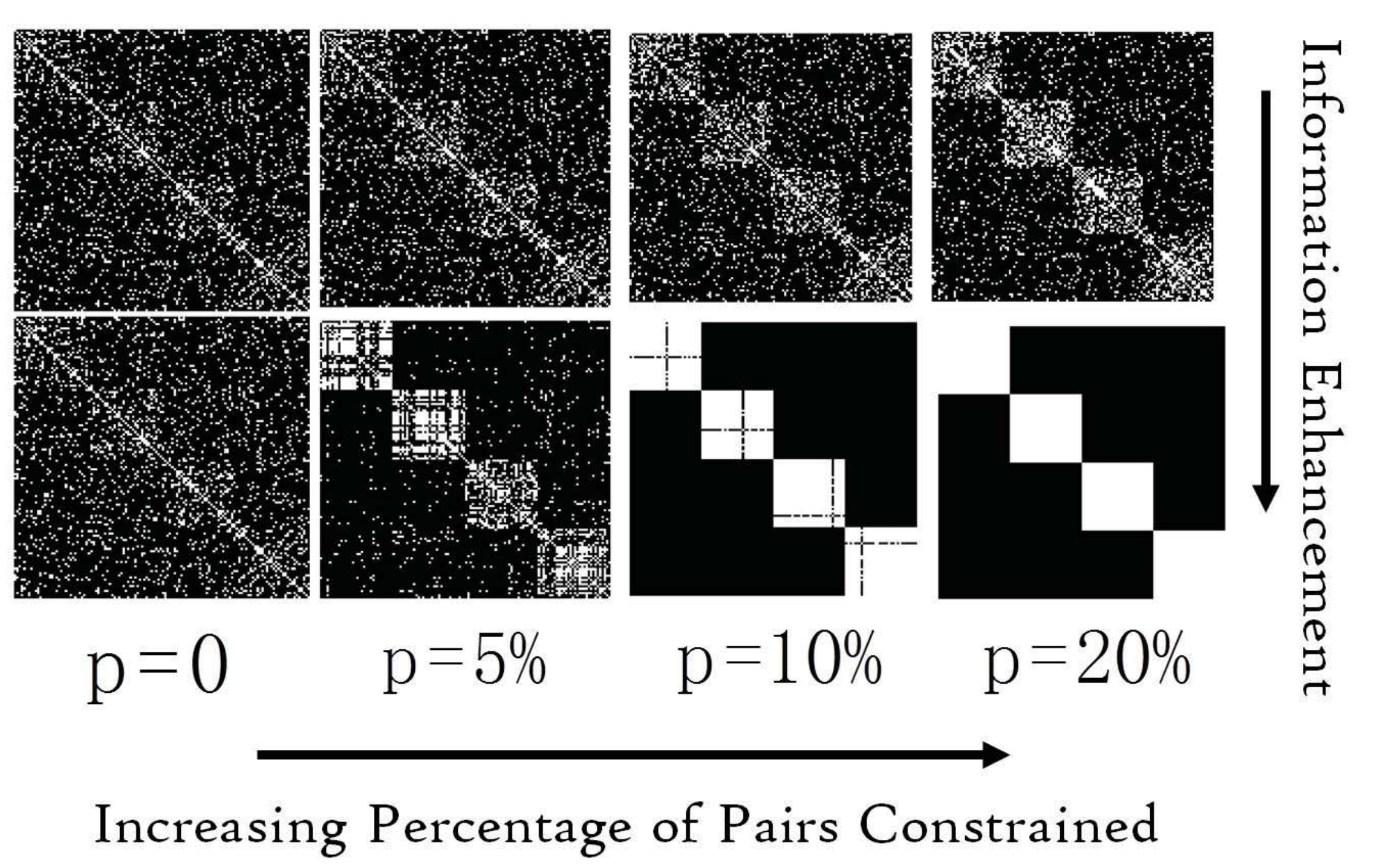}
 \caption{An illustrative example to show the effectiveness of information enhancement. $p$ is the percentage of pairs constrained. For the GN network, there are $128\times(128-1)/2=8128$ node pairs available.}\label{Fig:01}
\end{figure}

After incorporating the background information into the adjacency matrix, we can then apply some unsupervised learning models, such as nonnegative matrix factorization (NMF), spectral clustering and InfoMap, on them for community structure detection.

\subsection{Nonnegative Matrix Factorization (NMF)}
The model of NMF is often formulated as the following nonlinear programming \cite{nmf99,nmf00}:
\begin{eqnarray*}\label{eq:01}
\min_{W, H} & & \|X-WH\|_F^2\\
      s.t.           & & W, H\geqslant0,
\end{eqnarray*}
or in other words, given an nonnegative matrix $X$ of size $n\times n,$, we try to find two nonnegative matrices, $W$ of size $n\times k$ and $H$ of size $k\times n$, such that: $X\approx WH.$ The objective matrix $X$ for NMF can be selected as $A$, $B^{[1]},$ or $B^{[2]}$. The communities of the network can be revealed from $H$: node $i$ is of community $k$ if $H_{ki}$ is the largest element in the $i$th column.
The algorithm of multiplicative update rules for NMF can be summarized in Algorithm \ref{Al:01}.
\begin{algorithm}[h]
\caption{Nonnegative Matrix Factorization (Least Squares Error)}
\label{Al:01}
\begin{algorithmic}[1]
\REQUIRE $X,$ iter \ \ \% In this paper, the iteration number iter is set to 100.
\ENSURE $W, H.$
\FOR{$t=1:$iter}
\STATE \vspace{3mm}
$
\displaystyle W_{ik}:=W_{ik}\frac{(XH^T)_{ik}}{(WHH^T)_{ik}}
$
\STATE \vspace{3mm}
$
\displaystyle H_{ik}:=H_{ik}\frac{(W^TX)_{ik}}{(W^TWH)_{ik}}
$
\ENDFOR
\end{algorithmic}
\end{algorithm}

\subsection{Spectral Clustering}\label{spectral}
Spectral clustering is another powerful tool for unsupervised learning. The standard algorithm can be summarized in Algorithm \ref{Al:02} \cite{ng2002spectral}.

\begin{algorithm}
\caption{Spectral Clustering}
\label{Al:02}
\begin{algorithmic}[1]
\REQUIRE $B\in\mathbb{R}^{n\times n}$
\ENSURE Community Label $Y\in\mathbb{R}^{n\times 1}$ of the $n$ nodes
\STATE $L = D^{1/2}BD^{1/2}$, where $D$ is the diagonal matrix with the element $D_{ii}=\sum_jB_{ij}$ .
\STATE Forming the matrix $X=[x_1, x_2, \cdots, x_k]\in\mathbb{R}^{n\times k}$, where $x_i, i=1,2,\cdots,k$ are the top $k$ eigenvectors of $L$.
\STATE Normalizing $X$ so that rows of $X$ have the same $L_2$ norm: $X_{ij}=X_{ij}/(\sum_jX_{ij}^2)^{1/2}.$
\STATE Clustering rows of $X$ into $k$ clusters by K-means.
\STATE $Y_{i}=j$ if the $i$th row is assigned to cluster $j$.
\end{algorithmic}
\end{algorithm}
\subsection{InfoMap \cite{infomap}} This model grows out of information theory, and tries to detect the communities by minimizing the expected description length of random walks on the network. The model is among the most recommended approaches especially when there is no prior information on the network \cite{comparison}.
\section*{Acknowledgement}
This work is supported by the National Natural Science
Foundation of China under Grant No. 61203295.
\section*{References}

\end{document}